# The Influence of Crystallinity Degree on the Glycine Decomposition Induced by 1 MeV Proton Bombardment in Space Analog Conditions.

**Sergio Pilling**[1], **Luiz A. V. Mendes**[2], **Vinicius Bordalo**[3], **Christian F. M. Guaman**[3], **Cássia R. Ponciano**[3], **Enio F. da Silveira**[3]

[1] – Universidade do Vale do Paraíba (UNIVAP), São José dos Campos, SP, Brazil.

[2] – Universidade Federal da Bahia, (UFBA), Salvador, BA, Brazil.

[3] – Pontifícia Universidade Católica do Rio de Janeiro (PUC-Rio), Rio de Janeiro, RJ, Brazil.

e-mail: sergiopilling@pq.cnpq.br*;

**Abstract:**

Glycine is the simplest proteinaceous amino acid and is present in all life forms on Earth. In aqueous solutions, it appears mainly as zwitterion glycine ($^{+}NH_3CH_2COO^{-}$); however, in solid phase, it may be found in amorphous or the crystalline ($\alpha$, $\beta$, and $\gamma$) forms. The crystalline forms differ from each other by the packing of zwitterions in the unitary cells and by the number of intermolecular hydrogen bonds. This molecular species has been extensively detected in carbonaceous meteorites and was recently observed in the cometary samples returned to Earth by NASA's Stardust spacecraft. In space, glycine is exposed to several radiation fields at different temperatures. We present an experimental study on the destruction of zwitterionic glycine crystals at room temperature by 1 MeV protons, in which the dependence of the destruction rates of the $\alpha$-glycine and $\beta$-glycine crystals on bombardment fluence is investigated. The samples were analyzed in-situ by Fourier transformed infrared spectrometry at different proton fluences. The experiments occurred under an ultra-high vacuum conditions at the Van de Graaff accelerator's Lab at PUC-Rio, Brazil.

For low fluences, the dissociation cross section of $\alpha$-glycine was observed to be $2.5 \times 10^{-14}$ cm$^2$, a value roughly 5 times higher than the dissociation cross section found for $\beta$-glycine. The estimated half-lives of $\alpha$-glycine and $\beta$-glycine zwitterionic forms extrapolated to Earth orbit environment, are $9\times10^5$ and $4\times10^6$ years, respectively. In diffuse interstellar medium the estimated values are one order of magnitude lower. These results suggest that pristine interstellar $\beta$-glycine is the one most likely to survive the hostile environments of space radiation. A small feature around 1650-1700 cm$^{-1}$, tentatively attributed to amide functional group, was observed in the IR spectra of irradiated samples suggesting that cosmic rays may induce peptide bond synthesis in glycine crystals. Combining this finding with the fact that this form has the highest solubility among the other glycine polymorphs, we suggest that $\beta$-glycine is the one most likely to produce the first peptides on the primitive Earth.

**Keywords**: Astrobiology, Laboratory Investigation, Prebiotic Chemistry, Radiation Resistance, Interstellar Molecules

## 1. Introduction

The delivery of organic matter to the primitive Earth via comets and meteorites has long been hypothesized to be an important source for prebiotic compounds such as amino acids, nucleobases or precursors species that contributed to the development of prebiotic chemistry, which may led to the emergence of life on Earth (e.g. Oró, 1961; Chyba and Sagan, 1992; Aiello *et al.*, 2005). Significant evidence exists for an extraterrestrial origin, because observations have shown the presence of many organic molecules in interstellar medium (ISM), comets, and meteorites (e.g. Stoks and Schwartz, 1981; Cronin, 1998; Charnley *et al.* 2002; Glavin and Dworkin, 2009). More than 70 amino acids (including glycine) have been identified in meteorites, as in the Murchison meteorite (Cronin and Pizzarello, 1983; Glavin and Dworkin 2009; Glavin *et al.* 2011). Elsila *et al.* (2009), analyzed the cometary samples returned to Earth by NASA's Stardust spacecraft and found glycine amino acid among several other amine species.

In spite of the recurrent detection of amino acids in carbonaceous chondrites and in one comet, the observation of these molecules in the ISM has been a difficult task which have involved several unsuccessful attempts. For example, Kuan *et al.* (2003) searched for glycine ($NH_2CH_2COOH$), through hot molecular cores associated with the star-forming regions Sgr B2(N-LMH), Orion KL, and W51 e1/e2 and derived some upper limits for molecular abundances but these identifications have not yet been confirmed (Snyder *et al.* 2005; Cunningham *et al.* 2007). However, theoretical studies and laboratory experiments have shown that glycine is among the organic molecules that can be produced in space (e.g. Bernstein *et al.* 2002; Munoz-Caro *et al.* 2002; Nuevo *et al.* 2008; Nuevo *et al.* 2009; Pilling *et al.* 2010a, b; Nuevo *et al.* 2012).

Glycine is the simplest proteinaceous amino acid, the building blocks of proteins, and is an essential component of all living systems. In gas phase, glycine is observed in three conformers (I, II and III). Conformer I is the most abundant (~70-80%) and conformer II the most thermodynamically favorable (Ivanov *et al.* 1999). A conformer is an isomer of a molecule that differs from another isomer by the rotation of a single bond in the molecule.

In aqueous solution, glycine is found mostly as zwitterionic glycine ($^+NH_3CH_2COO^-$), although some glycine zwitterion clusters could also be present depending on their concentration. In the solid phase, glycine is crystallized with different structures (polymorphs): α-, β-, and γ-glycine forms. Figure 1 shows the most common glycine structures in different media (adapted from Ivanov *et al.* 1999; Fábián and Kálmán 2004; and Boldyreva 2008).

As discussed by Liu *et al.* (2008), the configuration of states of glycine zwitterions in these polymorphs differs from one another only by the angle between C-N and the least-square plane (a twist around the C-C bond). According to these authors, these forms are the consequences of intricate packing of molecules in the crystal lattice and the competition of several kinds of interactions: i) van der Waals, ii) electrostatic and iii) framework hydrogen bonding. In addition, the network of hydrogen bonding may plays an essential role in the organization of the crystal structure of glycine. The most favorable of these three polymorphs from the thermodynamic point of view is α-glycine (more hydrogen bonds between circumjacent zwitterions). Due to this strong intermolecular coupling, this species also has the highest density among the other polymorphs. However, as discussed by Bouchard *et al.* (2007), the solubility of β-glycine was significantly higher up to 17 % of the value attributed to α-glycine. Iitaka (1960) pointed out that the amount of glycine zwitterions in each crystal cell differs among the polymorphs, with 4 molecules in the α-form, 2 in the β-form, and 3 in the γ-form.

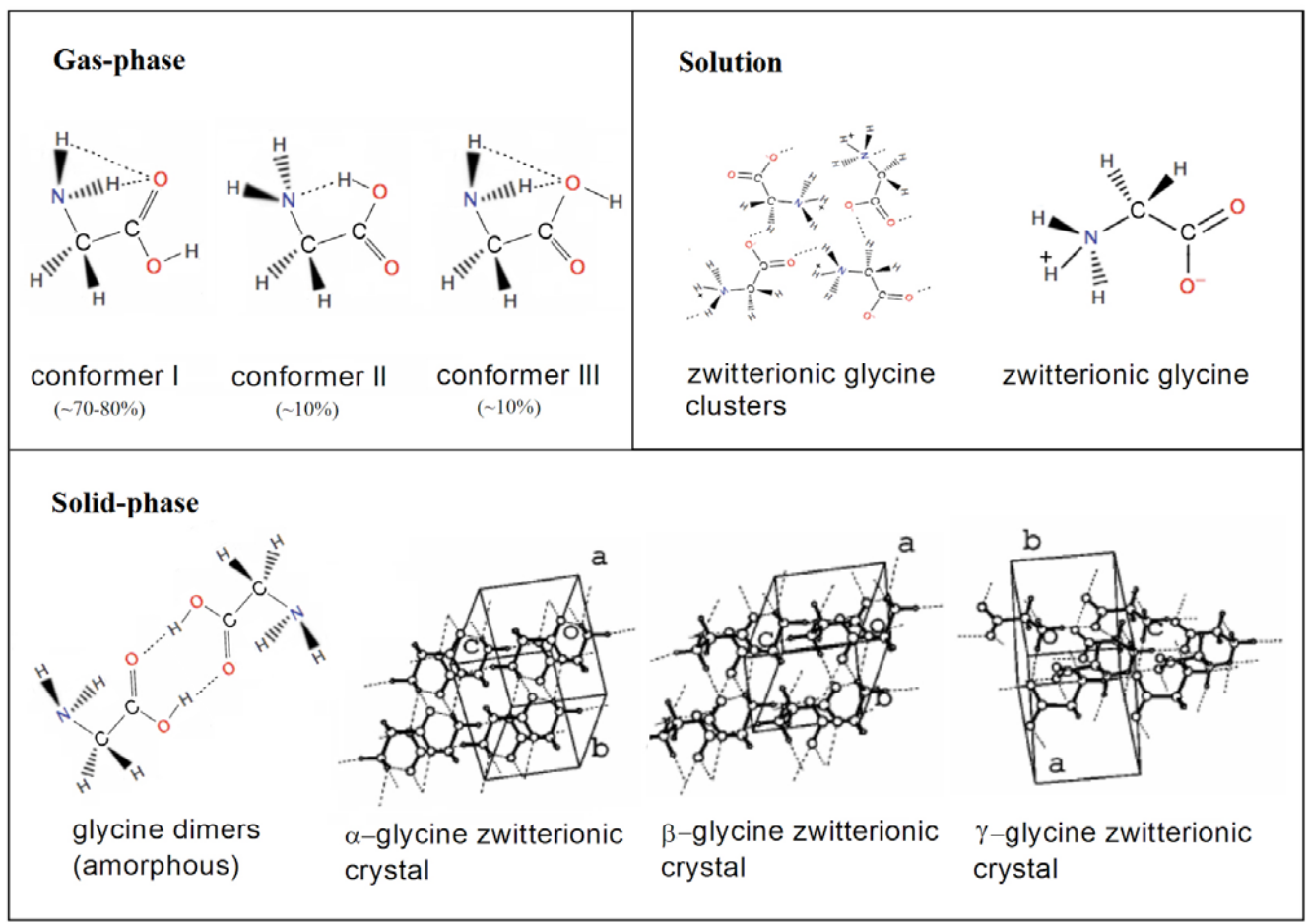

**Figure 1 - The most common glycine structures in different media (adapted from Ivanov *et al.* 1999; Fábián and Kálmán 2004; and Boldyreva 2008). The a, b, and c labels in the structure of the zwitterionic crystal indicate the axis of the unitary crystal cell.**

The formation of amorphous glycine was discussed by Maté *et al.* (2011). They observed that when gaseous glycine (neutral) adsorbs in non-polar frozen surfaces at 25 K (e.g. $CO_2$ and $CH_4$-rich ices) all glycine is in neutral forming dimers (see Fig. 1). For frozen polar surfaces at 25 K (e.g $H_2O$ rich ices), at least 50% of glycine is in neutral form. However, if the sample is heated to temperatures around 100 K, most of glycine molecules become zwitterionic. Following theses authors, virtually all glycine molecules are in the zwitterionic form at temperatures higher than 150 K. Lovas *et al.* (1995 and references therein) have pointed out that although glycine conformer II has the most intense lines in the theoretical radio spectra in respect to the other conformers, the most probable species in diffuse interstellar medium (DISM) should be glycine conformer I, due to its lowest energy among the conformers. Hypothetically speaking, if such species in the interstellar gas adsorb on warm (> 150 K) silicate/carbon interstellar grains, they would have enough energy to become glycine conformer II on the surfaces. Following Ivanov *et al.* (1999), such adsorbed conformers would be stabilized in the form of β-glycine zwitterionic crystal. However, if neutral glycine conformer I molecules adsorb on very cold interstellar grains (~ 20 K), for example inside molecular clouds, they would stabilize on the surface as glycine I dimers (amorphous). The authors also discussed that if cold grains are exposed to temperatures higher than 120-150 K, due to either by UV or soft-X ray heating promoted by a nearby young star in photodissociation regions or by a shock front of a nearby supernovae, the glycine dimers will be destroyed, which leads to the formation of β-glycine crystal. In protostellar or planetary system environment, β-glycine crystals may undergo phase transition by contact with liquid aqueous environment to become α-glycine crystals (see Glavin and Dworkin 2009; Glavin *et al.* 2011). This may happen, for example, inside a comet at local melting places induced by radioactive element heating. In addition, such β-glycine → α-glycine transformation may also occur in the subsurface aqueous environment of frozen icy moon/planetesimal (parent body) or in a moon/planet with liquid environments on the surface when the solvent evaporates. Such space environments, as well as the grains in ISM are permanently exposed to ionizing agents (photons, fast electrons and ions), which induce chemical reactions. The study of these chemical transformations is crucial to understand which molecules are the most likely to survive the extended voyage from ISM to a planet inside a habitable zone and participate (or trigger) the early steps towards the formation of life.

Radio observations suggest that the abundance of glycine in space (gas phase) must be very low (only upper limits were obtained). However, if we consider that small amount of glycine is present in ISM and that a small amount is continuously being produced from gas phase and deposited onto grains or produced directly on grain surfaces (e.g. Woon 2002; Zhu and Ho 2004; Pilling *et al.* 2011b), its abundance on grains will increase with time during the evolution of an interstellar or protostellar cloud. Moreover, if such grains were exposed to high temperatures, for example around 200 K, all the volatiles will desorb including water, leaving non-volatiles molecules such as glycine in the solid phase. Experimentally, glycine desorbs only at temperatures around 350-400 K under ultra-high vacuum conditions (e.g. Pilling *et al.* 2011a). If these grains undergo several heating and freezing cycles, such as the ones expected to occur around star-forming regions, the average number density of glycine (or other nonvolatile species) on grain surfaces will increase even more with time. When the number density reaches a certain value, the production of localized crystalline structures, such as β-glycine or α-glycine, cloud occur. Glycine crystals have about 20 molecules per cubic nanometer (Iitaka 1960).

The main goal of this study is to investigate the stability of different glycine polymorphs (α and β) irradiated by 1 MeV protons in simulated space conditions. Such glycine polymorph crystals are predicted to exist in icy space environments in which the temperature is higher than 100-120 K (for β-glycine) or in ices that undergo aqueous alteration (for α-glycine). The precise amount of interstellar glycine in gas phase is still unknown, and whether glycine is abundant enough to become crystal in space environments is still an open question. Future observations and sample extraction from comets or asteroids will help to clarify this issue. Therefore, until concrete evidence is found that glycine (free gas-phase molecule or crystals) is not present in space, investigation of the interaction of such important prebiotic species with space radiation analogs cannot be ruled out. Section 2 presents the experimental methodology employed. The obtained dissociation cross sections are presented in Section 3, followed by a discussion on astrophysical implications in Section 4. The conclusion and final remarks are in Section 5.

## 2 Experimental Setup

The Van de Graaff accelerator of the Pontifical Catholic University at Rio de Janeiro (PUC-Rio) was employed to provide 1 MeV H$^+$ beam. The projectiles traversed the sample, producing effects similar to those produced by energetic protons from solar wind and galactic cosmic rays. The choice of the energy of the beam was purely experimental. The energy distribution of protons in cosmic rays (Drury *et al.* 1999; Shen *et al.* 2004) or in the solar wind is very wide (e.g. Lin *et al*. 1974; Drury *et al.* 1999). Moore *et al.* (2001) pointed out that for solar wind, MeV protons dominate. For cosmic rays, the maximum occurs at hundreds of MeV. However, this energy range is very difficult to obtain in laboratory while in the energy range of few MeV is easy. Note that interactions involving 1 MeV protons and the very energetic ones, such as the 200 MeV protons, are of the same type, both occur with the electrons of the target (electronic stopping power regime). This imply that absolute numbers can change but the physical phenomena involved is the same.

The experiments were performed inside an ultra-high vacuum chamber (pressure around $1\times10^{-8}$ mbar) at room temperature (~ 300 K). In space, molecules undergo a very wide range of temperatures from 10 K to 10000 K (being fully dissociated). Our experiments simulate the temperature range in which water molecules are not present in solid phase but several refractory species (including glycine) are still present. Such a scenario occurs at the interface between molecular clouds and DISM, photodissociation regions, protoplanetary grains, and some surfaces of solar system bodies. Gerakines *et al.* (2012) recently investigated the effects of temperature on the dissociation of solid glycine by energetic protons. A comparison between our data and the data obtained at low temperature will be further discussed.

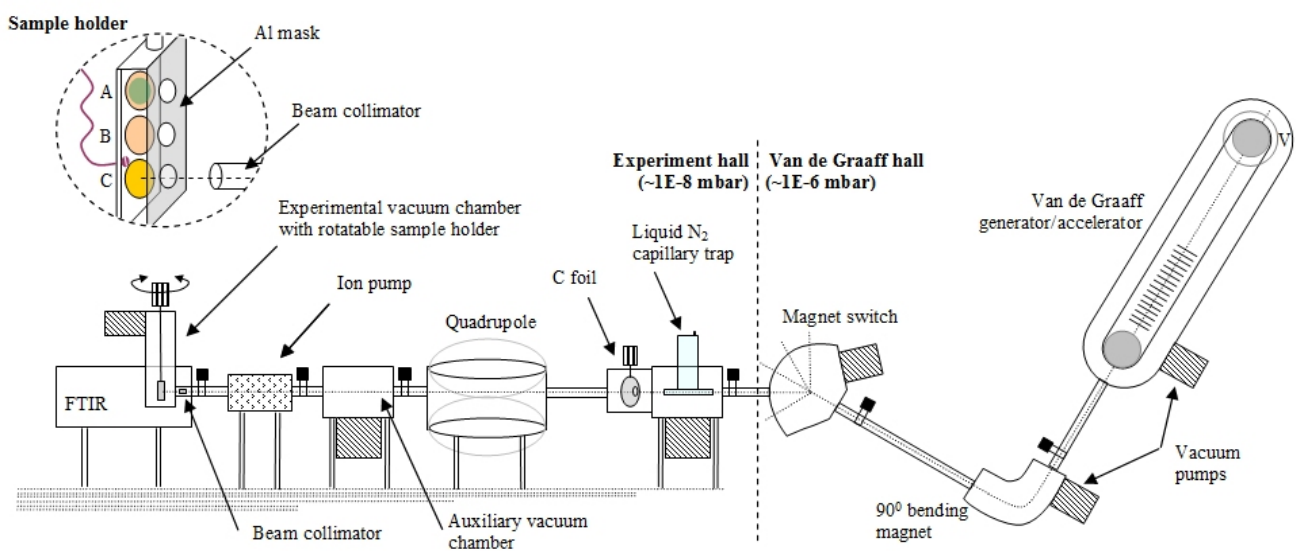

**Figure 2 - Schematic illustration of the experimental setup mounted at PUC-Rio Van de Graaff accelerator. The inset figure shows details of the sample holder (A: sample + substrate; B: clean substrate; C: Cu foil).**

The samples were analyzed in-situ by Fourier transformed infrared (FTIR) spectrometry at different proton fluences. Figure 2 presents a schematic illustration of the employed experimental setup. The inset figure shows details of the sample holder (A: sample + substrate; B: clean substrate; C: Cu foil). The carbon foil and the liquid $N_2$ capillary trap guaranteed the cleanness inside the experimental chamber. The carbon foil contributed to the bombardment homogeneity because the straggling of the proton beam increases the beam spot area. A quadrupole magnet along the beam line allowed a fine tuning of the beam optics.

A beam fluence calibration was performed before the sample irradiation. A copper foil connected to a microamperimeter was placed at the sample position C, around which a conducting circular collimator (~ 6 mm of inner diameter) is connected to a second microamperimeter (see inset Fig. 2). With the ion beam on, the ratio of both currents was measured. The sample holder was moved down, bringing the non-conducting sample (at position B) to the previous position of the copper foil. The beam flux (number of projectiles per square centimeter and second) on the target was then determined by multiplying the collected charge in the collimator by the ratio obtained previously. This ratio was checked periodically, before, during and after irradiation. A grounded aluminum mask with 3 holes of 5 mm diameter was placed just before the sample, to guarantee that all area probed by the FTIR was processed by the incoming ionizing radiation. A thin film of KBr was placed over the aluminum mask to allow good alignment of the target, the ionizing beam and the IR beam. In order to get a better charge measurement, a potential of +70 V was applied on the target holder and on the collimator to avoid secondary electrons from escaping. The sputtering of electrons from the target overvalues the current measurement. The estimated proton flux was around $2\text{-}3\times10^{10}$ ions cm$^{-2}$ s$^{-1}$.

The sample holder can be moved vertically for 3 positions (sample, a clean substrate, and Cu foil) and can be rotated allowing the irradiation (0º) and the FTIR analysis (90º) modes. A JASCO 4200 FTIR spectrometer was used to analyze the sample in the 3500-550 cm$^{-1}$ (2.8-18.2 μm) region with a spectral resolution of 1 cm$^{-1}$. Infrared (IR) spectra of clean substrates were acquired before each experiment for background correction purposes.

The production of glycine polymorphs in laboratory can be obtained in several ways. Briefly, α-glycine crystals can be produced, for example, by evaporation of aqueous glycine solution at room temperature, and β-glycine can be obtained from deposition of gas-phase glycine molecules (e.g. glycine conformer I) on warm substrates (> 150 K) inside vacuum. The β-glycine form is also called the metastable form of glycine, because it transforms spontaneously into α-glycine when the sample is exposed to moisture or grinding (Liu *et al.* 2008). The γ-glycine can be produced from evaporation of $NH_3$ rich aqueous solution containing glycine (Boldyreva 2008). Finally, in particular situations, glycine may also be found as amorphous solid. As pointed out by Ivanov *et al.* (1999), neutral glycine dimers can be produced by adsorption of glycine molecules from the gas phase (mainly glycine conformer I) on very cold (~ 20 K) surfaces under low pressure conditions. The formation of zwitterionic glycine in neutral glycine samples or glycine /water samples was studied theoretically by Ramaekers *et al.* (2004). They observed that a proton is transferred between the $NH_2$ and COOH groups of glycine when the amino acids is H-bounded with at least three water molecules or other glycine molecules. For a good review of glycine polymorphs see Boldyreva *et al.* (2003a; 2003b), Boldyreva (2008), and Liu *et al.* (2008).

In this paper, four samples were investigated: two produced by crystallization from the sublimation of glycine powder (99%, Sigma-Aldrich) inside an external evaporation vacuum chamber (β-glycine zwitterionic crystal), and two others produced by crystallization from drop casting of a glycine aqueous solution (0.1 M) dried by evaporation at room temperature (α-glycine zwitterionic crystal). For the drop-casting samples, a small amount of methanol was introduced into the glycine solution to decrease the superficial tension of the drop over the employed substrate (ZnSe). For the sublimated samples, an external evaporator chamber was employed to produce the thin films. Glycine powder was slowly heated up to 120-130 $^o$C for being deposited on two different substrates (KBr and ZnSe). After the samples were produced, they were kept inside a moisture free chamber until being introduced into the experimental chamber.

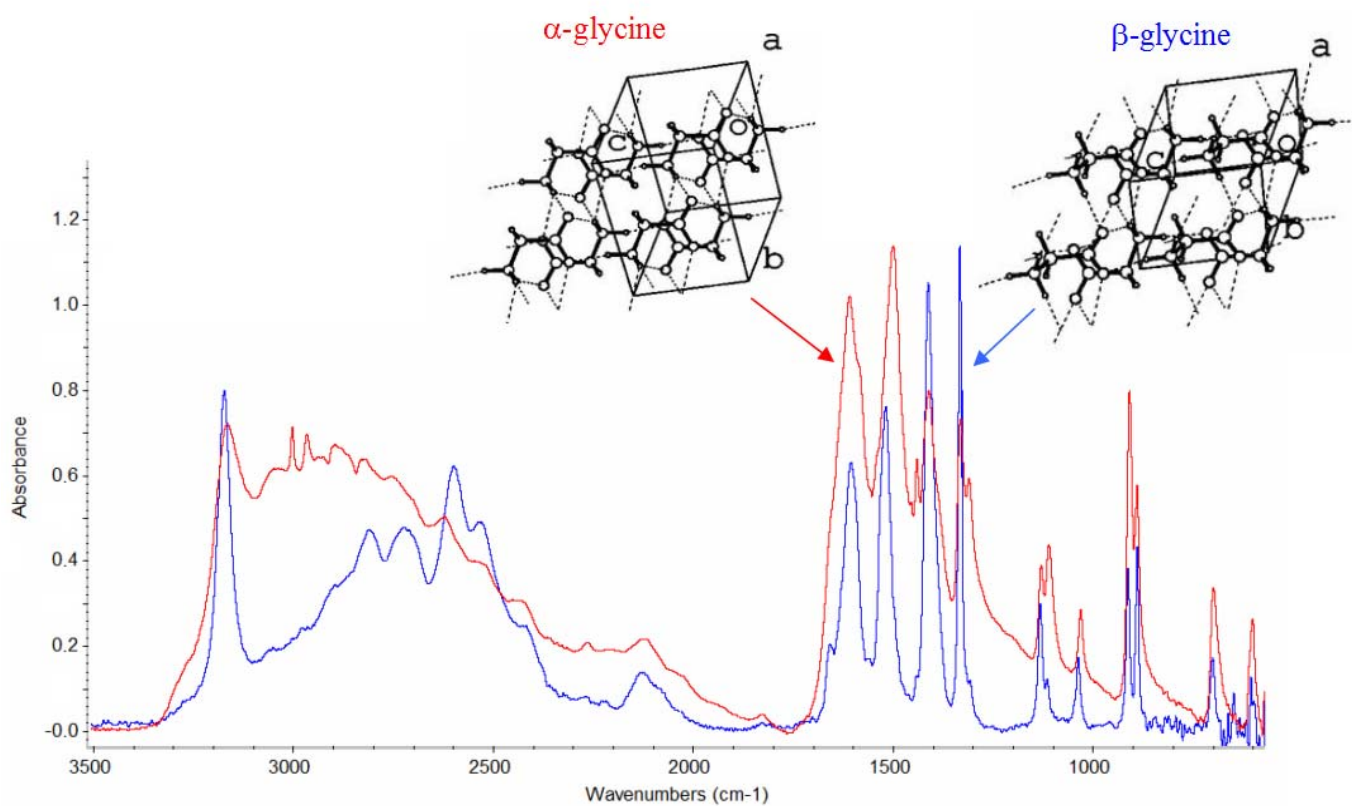

**Figure 3. Comparison between the IR spectra of two samples of glycine polymorphs: α-glycine zwitterionic crystal (obtained by evaporation of few drops of aqueous solution of glycine at 300 K) and β-glycine zwitterionic crystal (obtained by the deposition of sublimated glycine on substrate in vacuum). Each spectrum is the average of the spectra of two similar samples. Crystal structures of glycine polymorphs were taken from Fábián and Kálmán (2004). The a, b, and c labels indicate the axis the unitary cell in each crystal.**

Figure 3 shows a comparison between the IR spectra of the β-glycine and α-glycine samples at 300 K. Each spectrum is an average of the spectra of two similar samples. Crystal structures of glycine polymorphs were taken from Fábián and Kálmán (2004). The properties of the four samples investigated in this study are listed in Table 1.

**Table 1 - Glycine sample characteristics and initial parameters.**

| Experiment label | Sample crystallization from | Initial column density[c] ($10^{17}$ molecule cm$^{-2}$) | Thickness[c] (µm) | Mass[d] (µg) |
|---|---|---|---|---|
| β1 | gas phase on ZnSe substrate[a] | 3.0 | 0.3 | 7 |
| β2 | gas phase on KBr substrate[a] | 4.5 | 0.5 | 11 |
| α1 | solution on ZnSe substrate[b] | 9.2 | 1.0 | 22 |
| α2 | solution on ZnSe substrate[b] | 14.5 | 1.6 | 34 |

[a] performed in evaporation vacuum chamber from glycine powder sublimation around 120-130 °C, on substrate at room temperature.

[b] evaporation of few drops of 0.1M glycine aqueous solution with traces of methanol to decrease the surface tension of the drops.

[c] Using the Glycine IR band at 1034 cm$^{-1}$ (vCN vibration mode) with band strength of $1.38 \times 10^{-18}$ cm molec$^{-1}$ (Holton et al. 2005).

[d] The irradiation area is 0.19 cm$^2$.00

The initial column density, thickness, and mass of the samples were determined by employing the methodology described in Pilling *et al.* (2011a). The initial sample thicknesses were between 0.3 and 1.6 µm (Table 1). These values were determined by measuring in the IR spectra, the glycine peak area at 1034 cm$^{-1}$ (ν CN vibration mode), corresponding to a transition with a band strength of $1.38 \times 10^{-18}$ cm molecule$^{-1}$ (Holtom *et al.* 2005). In the current experiment, the location of the center of this band was observed to be 1039 cm$^{-1}$ (β-glycine) and 1034 cm$^{-1}$ (α-glycine). The sample density and its irradiated area were 1.16 g cm$^{-3}$ and 0.19 cm$^2$, respectively.

The amount of energy delivered by the projectile inside the glycine samples was calculated with the program SRIM - Stopping and Range of Ions in Matter (Ziegler *et al.* 2008). For 1 MeV proton beam, the main mechanism of energy transfer to glycine is by electronic excitation and the energy rate delivered (stopping power) is about 28 keV/ µm or $3 \times 10^{-14}$ eV/(molec/cm$^2$) or around 30 eV in each glycine monolayer (~$10^{15}$ molec/cm$^2$). This means that each proton traversing the sample slows down mainly by inelastic collisions with its electrons. The penetration range of 1 MeV proton in glycine (density of 1.16 g cm$^{-3}$) is about 22.6 µm. Therefore, in this experiment the employed projectile goes through the sample losing only 4% (in the thickest sample) of its energy, and for practical effects, the beam energy was considered to be constant inside the sample.

### 3 Results

Figure 4a presents the evolution of IR spectra of β-glycine (from β2 experiment) as function of proton fluencies (0, $1.2 \times 10^{13}$, $1.2 \times 10^{14}$, $3.2 \times 10^{14}$, and $1.3 \times 10^{15}$ ion cm$^{-2}$). Figure 4b shows the evolution of IR spectra of α-glycine (from α1 experiment) as function of proton fluencies (0, $1.4 \times 10^{13}$, $1.3 \times 10^{14}$, and $3.2 \times 10^{14}$ ion cm$^{-2}$). Both experiments were performed at room temperature. The absorption peaks around 2910, 2850, and 1550 cm$^{-1}$ are artifacts from background subtraction. In both figures, each spectrum has an offset for a better view. A small bump was observed around 1650-1700

cm$^{-1}$ in the IR spectra of both samples at higher fluences which may be attributed to the formation of peptide bonds by the appearing of amide functional group (-NHCO-) in the sample (Silverstein and Webster 1998). A peptide bond is a covalent chemical bond formed between two molecules when the carboxyl group of one molecule reacts with the amino group of the other molecule, causing the release of a water molecule. Therefore, the process is a dehydration synthesis reaction and usually occurs between amino acids (Morrison and Boyd 1992). Peptide bond formation was also observed in experiments employing microwave radiation in solid samples (Basso *et al.* 2009) and with ultraviolet in gaseous samples (Lee *et al.* 2011 and references therein).

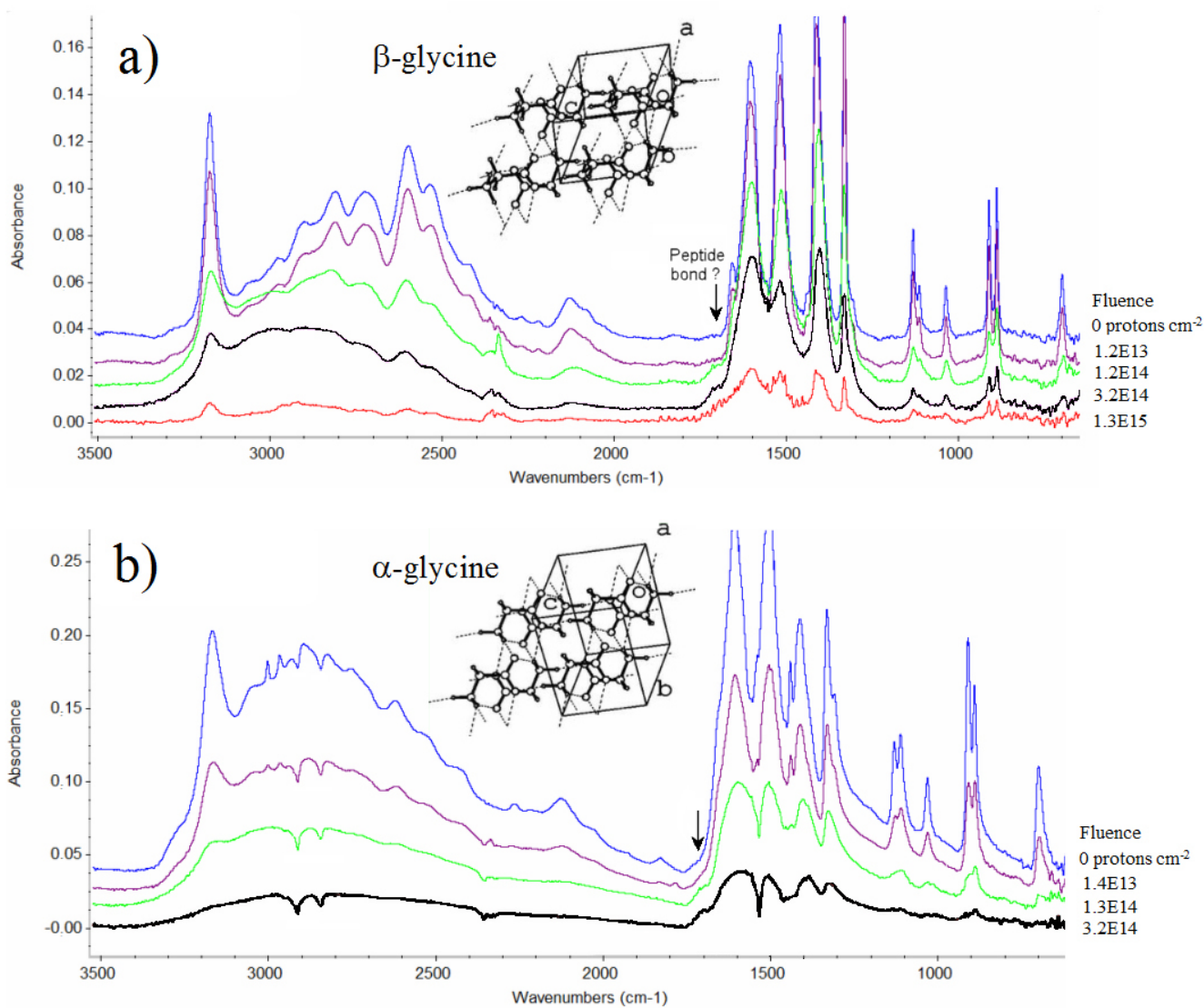

**Figure 4** - Evolution of IR spectra of: a) β-glycine (from β2 experiment) as function of proton fluence (0, $1.2 \times 10^{13}$, $1.2 \times 10^{14}$, $3.2 \times 10^{14}$, and $1.3 \times 10^{15}$ ion cm$^{-2}$); b) α-glycine (from α1 experiment) as function of proton fluence (0, $1.4 \times 10^{13}$, $1.3 \times 10^{14}$, and $3.2 \times 10^{14}$ ion cm$^{-2}$). The absorption peaks around 2910, 2850, and 1550 cm$^{-1}$ are artifacts from background subtraction. Crystal structures of glycine polymorphs were taken from Fábián and Kálmán (2004). Arrows indicate the position of amide functional group around 1650-1700 cm$^{-1}$ suggesting the formation of peptide bonds in the sample.

**Table 2** - Infrared absorption frequencies of α- and β-glycine zwitterionic crystals at the beginning of experiments (non-irradiated) and after $3.2\times10^{14}$ protons cm$^{-2}$ irradiation.

| Assigment[a] | β-glycine[b] | | α-glycine[c] | |
|---|---|---|---|---|
| | non irradiated | irradiated | non irradiated | irradiated |
| $\nu_{as}$ NH$_3^+$ | 3177 | 3175 | 3127 | 3172 |
| $\nu_{as}$NH$_3^+$ + $\nu$CN | 2600 | 2610 | 2622 | 2613 |
| $\nu_{as}$NH$_3^+$ + $\tau$NH$_3^+$ | 2130 | 2120 | 2128 | - |
| $\nu_{as}$ COO$^-$ | 1608 | 1603 | 1610 | 1591 |
| $\delta_{as}$ NH$_3^+$ | 1522 | 1522 | 1507 | 1507 |
| $\nu_s$ COO$^-$ | 1415 | 1406 | 1415 | 1388 |
| $\omega$ CH$_2$ | 1333 | 1333 | 1333 | 1325 |
| NH3 | 1134 | 1134 | 1133 | 1133 |
| $\rho$ NH$_3^+$ | 1117 | 1114 | 1114 | 1112 |
| $\nu$ CN | 1040 | 1038 | 1034 | 1034 |
| $\rho$ CH$_2$ | 914 | 914 | 912 | 917 |
| $\nu$ CC | 892 | 892 | 893 | 893 |
| $\delta$ COO | 703 | 700 | 700 | - |

[a] Holtom *et al.* (2005); Guan *et al.* (2010);

[b] from experiment β2 (see Table 1 and Figure 4a);

[c] from experiment α1 (see Table 1 and Figure 4b)

ν: stretching; δ: bending; ω: wagging; ρ: rocking; tw: twisting; τ: torsion; $_s$:symmetric; $_{as}$: asymmetric.

Comparing both experiments at one given fluence, for example at $3.2\times10^{14}$ ion cm$^{-2}$ (black curves in the Fig. 4), some peaks have disappeared in the spectrum of α-glycine, which suggests that this crystalline structure is the most sensitive to proton bombardment among the studied samples. Individual peaks, such as those around 3177 cm$^{-1}$ (attributed to N-H asymmetric stretching in the NH$_3^+$ group) and 1333 cm$^{-1}$ (wagging vibration mode of CH$_2$ group) have also shown more significant changes during the bombardment of α-glycine.

Table 2 lists the IR absorption frequencies of α- and β- glycine crystals at the beginning of experiments (non-irradiated) and after the fluence of $3.2 \times 10^{14}$ protons cm$^{-2}$. For both α- and β-glycine experiments, it was observed that after proton bombardment, several peaks were shifted slightly towards lower wavenumbers. This behavior becomes more prominent when α-glycine is bombarded.

### 4 Discussion

The evolution of the normalized area of selected IR bands of α- and β-glycine as function of the proton fluence **is** presented in Figure 5. The selected glycine bands are centered at: a) 3170 cm$^{-1}$ ($\nu_{as}$ NH$_3$), b) 1596 cm$^{-1}$ ($\nu_{as}$ COO), c) 1330 cm$^{-1}$ (w CH$_2$), d) 1034 cm$^{-1}$ ($\nu$ CN), e) 893 cm$^{-1}$ ($\nu$ CC), and f) integrated area from 3500 to 850 cm$^{-1}$. The hatched site over zwitterionic glycine structural formulae shows the chemical bonds studied in each panel. The obtained values of the dissociation cross sections ($\sigma_d$) are indicated. The molecular dissociation cross section considering a specific bond (or functional group) rupture was obtained by the Lambert-Beer´s related expression

$$\ln (A/A_o) = - \sigma_d \times F \qquad [1]$$

where A and Ao are the areas of the IR bands associated with a specific molecular vibration at a given fluence and at the beginning of experiment, respectively. F is the ion fluence in units of ion cm$^{-2}$. The values of molecular dissociation cross section considering specific bonds in α- and β-glycine crystals during 1 MeV proton bombardment are listed in Table 3.

The results displayed in Fig. 5 show that the glycine bonds present different fragility under irradiation by 1 MeV protons. Particularly, for β-glycine, the N-H and C-N bonds are much fragile when compared with those of the carboxylate group (COO$^-$). This fact can be quantified by the values of the dissociation cross sections, which are 5.6 x10$^{-15}$, 5.4 x10$^{-15}$ and 1.1x10$^{-15}$ cm$^{-2}$, respectively, as shown in Table 3. In this work, we consider that the dissociation cross section of glycine is mainly governed by the stability of the C-N bond (ν CN vibration mode at 1034 cm$^{-1}$), since the rupture of such bond will dismantle the entire molecule releasing both the amino and the carboxylic acid groups. In addition, the evolution ν CN vibration mode as function of fluence, presented the most similar behavior with Lambert-Beer´s law (a straight line) among the other molecular bonds (see Fig. 5) for α-glycine. Our results show that the dissociation cross section, based on the stability of the C-N bond, of α-glycine by 1 MeV protons is found to be roughly 5 times higher than the value obtained for β-glycine in the same experimental conditions. The values obtained for β-glycine and α-glycine were 5.4×10$^{-15}$ and 2.5×10$^{-14}$ cm$^2$, respectively.

For both studied glycine polymorphs, the molecular bond most sensitive to the incoming 1 MeV protons is the N-H bond, followed by the C-N bond. The dissociation cross section of α-glycine, considering only the N-H vibration mode, was 3.1x10$^{-14}$ cm$^2$, a value roughly 6 times higher than that determined for the same vibration mode of β-glycine. However, if the whole IR spectrum (850-3500 cm$^{-1}$) is considered, this difference is still higher, causing the dissociation cross section of α-glycine to be roughly 12 times higher than that for the β-glycine. The molecular bond less sensitive to the ionizing protons for α-glycine is the C-C bond (893 cm$^{-1}$), resulting a dissociation cross section of 1.6×10$^{-14}$ cm$^2$. Curiously, the least sensitive bond in β-glycine crystal is one of the most sensitive bond in α-glycine, the C-O in the carboxylate group (COO$^-$). The dissociation cross section, considering only this IR band in α-glycine, was measured to be 2.3×10$^{-14}$ cm$^2$, a value 21 times higher than that measured for β-glycine. Such behavior is probably associated with the packing of molecules in the crystal lattice. As discussed by Liu *et al.* (2008), the crystal structure of β-glycine consists of hydrogen-bonded monomers units, which is different from that of α-glycine formed by cyclic hydrogen-bounded pairs.

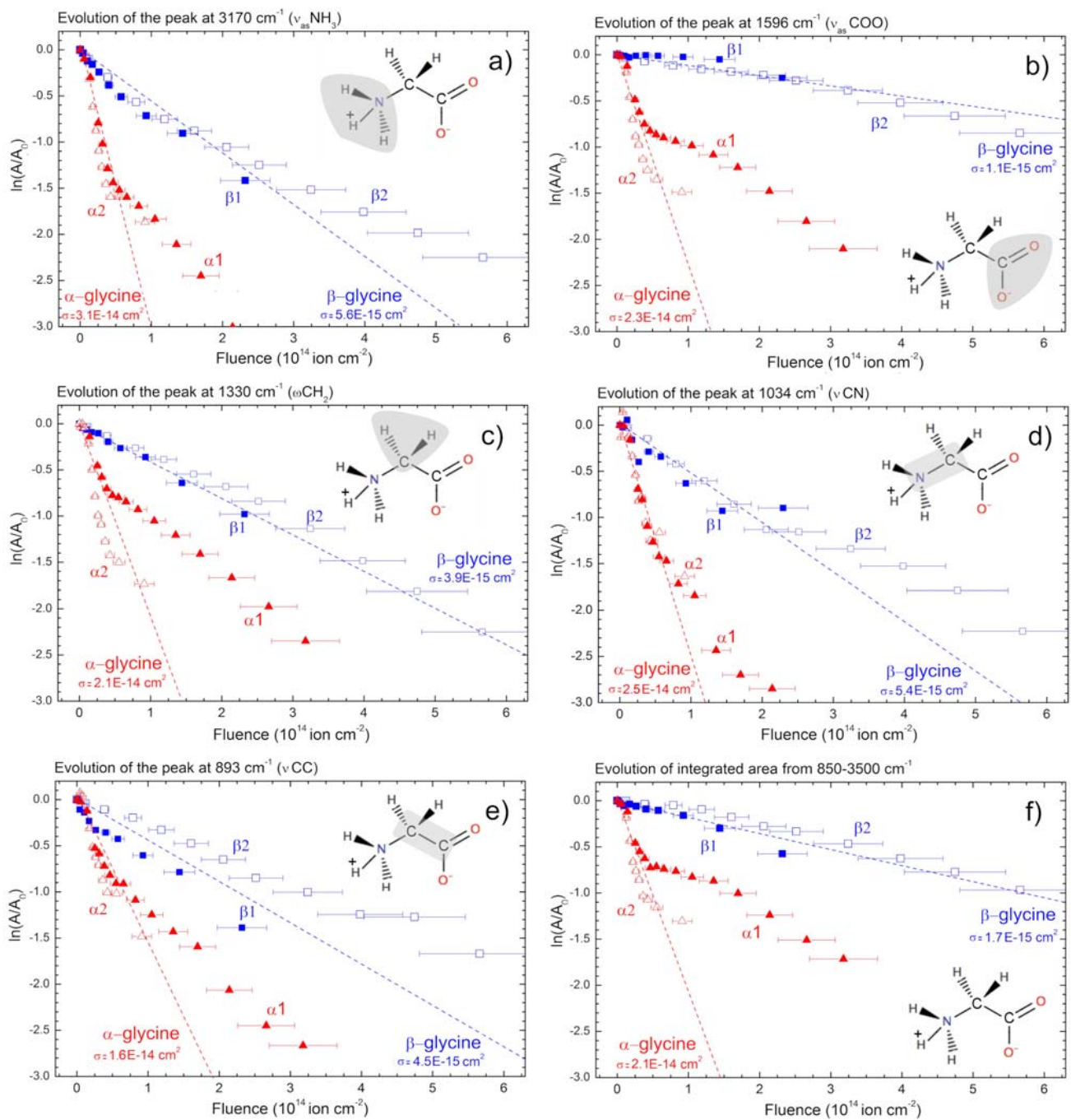

**Figure 5** - Evolution of the normalized peak areas as function of the proton fluence. The selected glycine IR bands are: a) 3170 cm$^{-1}$ ($\nu_{as}$ NH$_3$), b) 1596 cm$^{-1}$ ($\nu_{as}$ COO$^-$), c) 1330 cm$^{-1}$ (w CH$_2$), d) 1034 cm$^{-1}$ ($\nu$ CN), e) 893 cm$^{-1}$ ($\nu$ CC), and f) whole molecule using integrated area of IR spectra from 850 to 3500 cm$^{-1}$. The hatched site over zwitterionic glycine structural formulae shows the chemical bonds studied in each figure. The obtained values of the dissociation cross sections considering specific molecular bonds ($\sigma_d$) are indicated.

From our results, we suggest that, because α-glycine has the highest density among the glycine polymorphs, the number of molecules processed by each incoming ion is also highest and, therefore, resulted in the highest dissociation cross section observed. In addition, we observed that all molecular bonds in α-glycine crystal presented lower stability to 1 MeV protons when compared with β-glycine. In addition to the density, such behavior is probably also related with the framework hydrogen bonding in the crystal.

Other experimental studies on the stability of amino acids against radiation have considered the rupture of different bonds to quantify the amino acid destruction, such as the CO bond observed by the $COO^-$ symmetric stretch near 1400 cm$^{-1}$ (e.g. Gerakines *et al.* 2012), the carbonyl (C=O) stretching mode near 1780 cm$^{-1}$ (e.g. Ehrenfreund *et al.* 2001) or also by the combination of different bonds from the integrated absorbance in the IR spectra (e.g. ten Kate *et al.* 2005; Pilling *et al.* 2011a). Such experiments have obtained different values for dissociation cross section, even when they have been performed at similar condition. The choice of different molecular bond (IR feature) to characterize the molecular dissociation is an explanation for that issue.

Recently, Gerakines et al. (2012) have performed another set of experiments on solid glycine. The authors analyzed the stability of glycine in the presence or absence of water at low temperature (14, 100 and 140 K) during the bombardment with 0.8 MeV protons. The dissociation cross section obtained by these authors for β-glycine samples at 140 K was $1.2 \times 10^{-15}$ cm$^2$, which is virtually the same value than the obtained here at room temperature (only 10% higher), when we taking into account only the $COO^-$ group.

**Table 3 - Dissociation cross section ($\sigma_d$) for specific bonds of α- and β-glycine crystals submitted to 1 MeV proton bombardment.**

| Specific bonds (vibration mode) | $\sigma_d$ (β-glycine) (cm$^2$) | $\sigma_d$ (α-glycine) (cm$^2$) | Ratio[$\sigma_d$ (α-glycine) / $\sigma_d$ (β-glycine)] |
|---|---|---|---|
| N-H ($\nu_{as}$ NH$_3^+$) | $5.6 \times 10^{-15}$ | $3.1 \times 10^{-14}$ | 5.6 |
| C-N ($\nu$ CN) | $5.4 \times 10^{-15}$ | $2.5 \times 10^{-14}$ | 4.6 |
| C-C ($\nu$ CC) | $4.5 \times 10^{-15}$ | $1.6 \times 10^{-14}$ | 3.6 |
| C-H ($\omega$ CH$_2$) | $3.9 \times 10^{-15}$ | $2.1 \times 10^{-14}$ | 5.4 |
| C-O ($\nu_{as}$ COO$^-$) | $1.1 \times 10^{-15}$ | $2.3 \times 10^{-14}$ | 21 |
| Whole molecule (from 850 to 3500 cm$^{-1}$) | $1.7 \times 10^{-15}$ | $2.1 \times 10^{-14}$ | 12 |

**5- Astrophysical implication**

The stability of glycine for different ionizing agents in an attempt to investigate its survival in space environment simulations have been examined by several researchers in the last decade. These investigations have involved ultraviolet radiation (e.g., Ehrenfreund *et al.* 2001; ten Kate *et al.* 2005; Peeters *et al.* 2003; Guan *et al.* 2010; Ferreira-Rodrigues *et al.* 2011), X-rays (e.g., Pilling *et al.*

2011a), electron beams (e.g Abdoul-Carime and Sanche 2004) and fast ions (this study; Portugal *et al.* 2012, Gerakines et al. 2012). A list of dissociation cross sections ($\sigma_d$) taken from these studies, when available, is present in Table 4.

**Table 4 - Dissociation cross section ($\sigma_d$) and half-lives ($\tau_{1/2}$) of glycine polymorphs (and in gas phase) predicted for space environments exposed to different ionizing agents.**

| Ionizing agent | Sample properties | $\sigma_d$ (cm$^2$) | $\tau_{1/2}$ (Yr) Earth orbit | $\tau_{1/2}$ (Yr) DISM | $\tau_{1/2}$ (Yr) Dense Clouds | Ref |
|---|---|---|---|---|---|---|
| UV lamp (10 eV) | α/γ-glycine (300 K)[e] | - | $2.2 \times 10^{-2}$ | $2.2 \times 10^{3}$ | $2.0 \times 10^{8}$ | [1] |
| | Neutral glycine in argon matrix (12 K) | $1.2 \times 10^{-17}$ | $6.1 \times 10^{-5}$ | $1.8 \times 10^{1}$ | $1.8 \times 10^{6}$ | [3] |
| | Neutral glycine in argon matrix (12 K) | - | - | $3.2 \times 10^{2}$ | $3.2 \times 10^{7}$ | [9] |
| | Neutral glycine in nitrogen matrix (12 K) | - | - | $3.2 \times 10^{2}$ | $3.2 \times 10^{7}$ | [9] |
| | Neutral glycine in water matrix (12 K)[f] | - | - | $2.2 \times 10^{2}$ | $2.2 \times 10^{7}$ | [9] |
| | β- glycine (300 K)[a] | $2.4 \times 10^{-19}$ | $3.1 \times 10^{-1}$ | $9.2 \times 10^{2}$ | $9.2 \times 10^{7}$ | [6] |
| | α-glycine (300 K) positively charged[d] | $1.7 \times 10^{-16}$ | $4.3 \times 10^{-4}$ | $1.3 \times 10^{0}$ | $1.3 \times 10^{5}$ | [7] |
| | α-glycine (300 K) negatively charged[d] | $9.0 \times 10^{-18}$ | $8.1 \times 10^{-3}$ | $2.4 \times 10^{1}$ | $2.4 \times 10^{6}$ | [7] |
| Space UV | α/γ-glycine (250/300 K)[e] | - | $1.1 \times 10^{-2}$ | - | - | [1] |
| Soft X-ray (150 eV) | α-glycine (300 K)[d] | $4.0 \times 10^{-17}$ | $5.5 \times 10^{1}$ | $1.8 \times 10^{-2}$ | $1.8 \times 10^{2}$ | [2] |
| | Neutral glycine in gas phase (420 K) | $2.6 \times 10^{-18}$ | $8.4 \times 10^{2}$ | $2.8 \times 10^{-1}$ | $2.8 \times 10^{3}$ | [2] |
| Proton (0.8 MeV) | α-glycine (15 K)[c] | $2.9 \times 10^{-15}$ | $7.6 \times 10^{6}$ | $7.6 \times 10^{5}$ | $7.6 \times 10^{6}$ | [8] |
| | α-glycine (140K)[c] | $2.1 \times 10^{-15}$ | $1.0 \times 10^{7}$ | $1.0 \times 10^{6}$ | $1.0 \times 10^{7}$ | [8] |
| | Neutral glycine (15K)[b] | $2.2 \times 10^{-15}$ | $1.0 \times 10^{7}$ | $1.0 \times 10^{6}$ | $1.0 \times 10^{7}$ | [8] |
| | β-glycine (140K)[a] | $1.2 \times 10^{-15}$ | $1.8 \times 10^{7}$ | $1.8 \times 10^{6}$ | $1.8 \times 10^{7}$ | [8] |
| Proton (1 MeV) | α-glycine (300 K)[d] | $2.5 \times 10^{-14}$ | $8.8 \times 10^{5}$ | $8.8 \times 10^{4}$ | $8.8 \times 10^{5}$ | [4] |
| | β-glycine (300 K)[a] | $5.4 \times 10^{-15}$ | $4.1 \times 10^{6}$ | $4.1 \times 10^{5}$ | $4.1 \times 10^{6}$ | [4] |
| Nickel (46 MeV) | α-glycine (300 K)[d] | $1.3 \times 10^{-13}$ | $8.0 \times 10^{6}$ | $3.0 \times 10^{6}$ | $> 3 \times 10^{6}$ | [5] |
| | α-glycine (14 K)[d] | $\sim 3 \times 10^{-12}$ | $\sim 4 \times 10^{5}$ | $\sim 1 \times 10^{5}$ | $> 1 \times 10^{5}$ | [5] |

[a]Sublimated glycine inside vacuum chamber. [b]Glycine dimers. [c]Sublimated glycine inside vacuum chamber together with water. [d] Glycine film produced by dropped glycine aqueous solution. [e]From a film of β-glycine exposed to moisture. [f] Ten molecules o water for each glycine molecule.

[1] Guan et al (2010); [2] Pilling et al. (2011a); [3] Peeters et al. (2003); [4] This work; [5] Portugal et al. (2012); [6] ten Kate et al. (2005); [7] Ferreira-Rodrigues et al. (2011); [8] Gerakines et al. (2012), [9] Ehrenfreund et al. (2001).

Table 4 also presents the estimated half-lives ($\tau_{1/2}$) of glycine in space environments such as around Earth, in DSIM and in dense clouds in the ISM which are exposed to different radiation fields. In this table, excepting the work of the Guan *et al.* (2010) and Ehrenfreund *et al.* (2001), the half-lives were obtained directly from the expression

$$\tau_{1/2} = \ln 2 \, (\phi \, \sigma_d)^{-1} \qquad [2]$$

where ϕ is the ionizing agent flux in particles cm$^{-2}$ s$^{-1}$ and σ$_d$ is the dissociation cross section in cm$^2$. This equation does not depend on the molecular number density. For UV photons (~10 eV), the following fluxes were considered: ~3x10$^{11}$ photons cm$^{-2}$ s$^{-1}$ at Earth orbit (Huebner and Carpenter, 1979), ~10$^8$ photons cm$^{-2}$ s$^{-1}$ at DISM (Moore *et al.* 2001) and ~10$^3$ photons cm$^{-2}$ s$^{-1}$ at dense clouds (cosmic ray-induced UV flux; Prasad and Tarafdar 1983). For soft X-rays (~150 eV), the values adopted were: ~10$^7$ photons cm$^{-2}$ s$^{-1}$ at Earth orbit (Gueymard 2004), ~ 3 x 10$^{10}$ cm$^{-2}$ s$^{-1}$ photons cm$^{-2}$ s$^{-1}$ at DISM (estimated value for X-ray dissociated region, Pilling *et al.* 2011a), and 3x10$^6$ photons cm$^{-2}$ s$^{-1}$ at dense clouds (Pilling *et al.* 2011a). For the proton (1 MeV) component of cosmic rays, the following fluxes were considered: ~ 1 proton cm$^{-2}$ s$^{-1}$ at Earth Orbit (Lin *et al.* 1974), ~10 protons cm$^{-2}$ s$^{-1}$ at DISM (Strazzulla and Johnson, 1991), and ~1 proton cm$^{-2}$ s$^{-1}$ at dense clouds (Moore *et al.* 2001). Finally, for heavy component of cosmic rays (e.g. 46 MeV Ni) the values considered were: ~2 x 10$^{-2}$ ions cm$^{-2}$ s$^{-1}$ at Earth orbit and ~5x10$^{-2}$ ions cm$^{-2}$ s$^{-1}$ at DISM (Pilling *et al.* 2010a, b). The heavy ion flux inside dense clouds was estimated to be < 5x10$^{-2}$ ions cm$^{-2}$ s$^{-1}$.

As discussed in Section 4, the dissociation cross section of α-glycine by 1 MeV protons is roughly 5 times higher (considering only the disruption of C-N bond) than the value obtained for β-glycine in the same experimental conditions. A similar result was observed by comparing the data obtained by ten Kate *et al.* (2005) with the one from Ferreira-Rodrigues *et al.* (2011), for samples irradiated by UV photons. Gerakines *et al.* (2012), in another study performed with energetic protons (0.8 MeV), also observed that at the temperature of 140 K, the destruction of α-glycine by protons is approximately 2 times higher than that observed for β-glycine (produced after the heating of neutral glycine dimers initially at 15 K). In addition they observed that, for α-glycine, the lower the temperature the higher the dissociation cross section. Such temperature dependence on dissociation cross section was also found by Portugal *et al.* (2012) by bombarding glycine with swift heavy ions simulating the heavy component of cosmic rays. The result suggests that the pristine interstellar β-glycine form is the strongest glycine polymorph; therefore, the most able to survive the hostile space radiation environments.

Table 4 shows that soft X-rays are mainly responsible for shortening the half-life of glycine in dense clouds or, in other words, the glycine destruction in dense clouds is ruled by soft X-rays. This is also true for DISM, but with less intensity. However at Earth orbit, the dominant ionizing agent in the glycine dissociation is the UV photons. The effect of cosmic rays (light or heavy ions) in the glycine dissociation in dense clouds is roughly the same as that produced by UV photon in such media.

In space, glycine can be formed directed from gas-phase reactions (glycine conformers) or from surface/bulk induced reactions (β-glycine and α-glycine zwitterionic crystals or amorphous solid, depending on the temperature). Due to the low temperature of molecular clouds, glycine is expected to occur as neutral dimers, becoming an amorphous solid. However, if the ices are heated, glycine will transform into β-glycine crystals, and in the presence of water molecules, it will be transformed into α-glycine crystals. Figure 6 shows a scheme of the transformations of glycine crystalline structures in space after its formation from some gas-phase reactions and/or surface reactions (adapted from Liu *et al.* 2008 and Pilling *et al.* 2011). Recently, Glavin *et al.* 2011, in an experimental study on carbonaceous meteorites, found evidence of aqueous alteration of amino acids in solar system parent bodies, which suggests that α-glycine form may be dominant in such environments. Due to radioactive-induced melting inside comets, α-glycine can also be considered the most probable among the glycine polymorphs in such objects. Therefore, the study of stability of both α- and β-glycine forms under ionizing agents (analogous to the one found in space environments) is highly required to understand the presence of such compounds in the early Earth and in other planets with astrobiological potential.

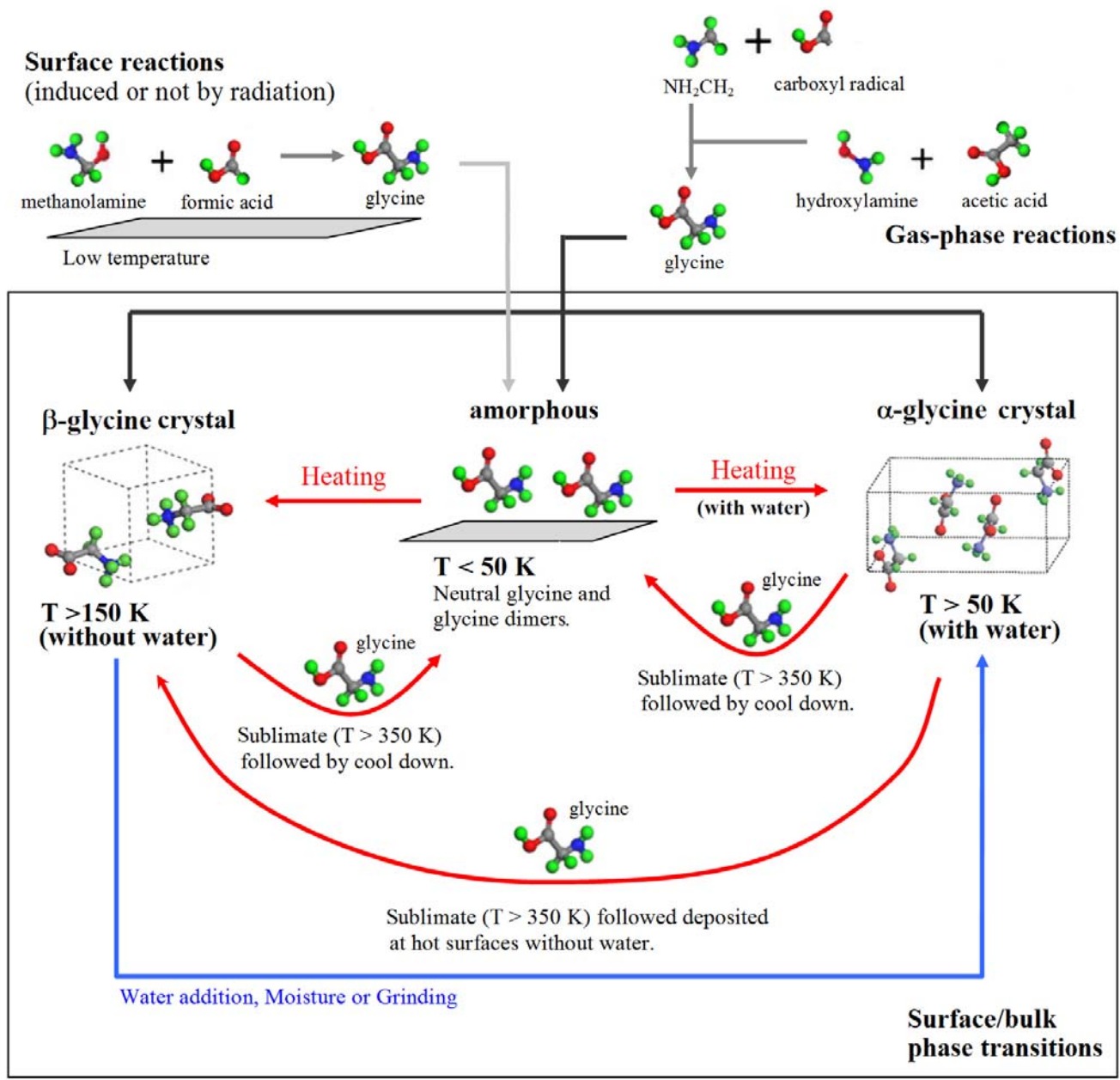

**Figure 6 - Scheme of the transformations of glycine crystalline structures in space after its formation from some gas-phase reactions and/or surface reactions (adapted from Liu *et al.* 2008 and Pilling *et al.* 2011b).**

Another important astrophysical implication of the present work is the possible peptide synthesis induced by cosmic ray analogs in solid amino acids. A small bump was observed around 1650-1700 cm$^{-1}$ in the IR spectra of both samples at higher proton fluences, which was tentatively attributed to the formation of amide functional group from glycine molecules, and suggesting a possible formation of a peptide bond with the release of water molecule. Peptides, such as di-glycine peptide (NH₂CH₂CONHCH₂COOH), have been detected in the Murchison and Yamato-791198 meteorites in low quantities (Shimoyama and Ogasawara 2002). Other linear and cyclic aliphatic amides, including monocarboxylic acid amides, dicarboxylic acid monoamides, N-acetyl amino acids, and others, have also been identified in the carbonaceous chondrites meteorites (e.g. Cooper and Cronin 1995). Comparing these detections with the possible observation of peptide bond in our work, we suggest that at least a fraction of peptides present in the early Earth may have had an extraterrestrial origin, possibly being induced by cosmic rays on grains in the ISM or in the solar protoplanetary disk. Recently, Halfen *et al.* (2011) have performed radio observations to detect and quantify peptide bonds in the ISM. They identified several lines of small amides toward the star-forming region Sgr B2 (N), including acetamide ($CH_3CONH_2$) and formamide ($NH_2CHO$), and pointed out that such species may participate of an alternative scenario for the formation of larger peptide molecules, as opposed to amino acids.

Recent experiment performed by our group on the processing of α-glycine by 46 MeV Nickel ions, also under ultra-high vacuum conditions, have shown the formation of water during the bombardment at 14 K (Portugal et al. 2012). Such behavior may be explained by the water releasing from glycine molecules during the formation of peptide bond. Future experiments focusing on the production of amides formed by the bombardments of amino acid in space analog conditions may help to clarify this issue.

## 6. Conclusions

We performed an experimental study to simulate the effect of energetic proton (1 MeV) bombardment on two types of glycine crystals (α- and β-glycine zwitterionic polymorphs). The experiment simulates the interaction of typical interstellar glycine molecule in space environment such as comets/asteroids/moon surfaces and interstellar grains with galactic cosmic rays and energetic protons from solar wind.

For low proton fluences, the dissociation cross section of α-glycine is $2.5 \times 10^{14}$ cm², a value roughly 5 times higher than the dissociation cross section found for β-glycine. Comparing this result with others from literature, we concluded that α-glycine also has higher sensitivity to UV photons than β-glycine. Curiously, although β-glycine is the less favorable from a thermodynamic point of view and has the smallest density among the other glycine polymorphs (Pervolich et al. 2001), it is more resistant to ionizing agents than α-glycine. Following Liu et al. (2008), the β-glycine form has the weakest hydrogen-bond networks. We suggest that its lower density in comparison with α-glycine was responsible for its greater stability against ionizing agents.

The estimated half-lives of α-glycine and β-glycine zwitterionic forms bombarded by 1 MeV protons at room temperature, extrapolated to Earth orbit environment, are $9 \times 10^5$ and $4 \times 10^6$ years, respectively. In DISM, the estimated values are one order of magnitude lower. We also observed that the effect of cosmic rays (light or heavy ions) in the glycine dissociation in dense clouds is roughly the same as the UV photon filed in such media. In dense clouds, soft X-rays are mainly responsible for shortening the lifetime of glycine species. This is also true for DISM but with less intensity. However, for the interplanetary medium (for example at Earth orbit) the dominant ionizing agent in the glycine dissociation is the UV photons.

The results suggest that the pristine interstellar β-glycine form should be the most probable kind of glycine polymorph to survive the hostile space radiation environments (considering both UV and protons). In addition, a small feature around 1650-1700 cm$^{-1}$, tentatively attributed to amide functional group, was observed in the IR spectra of irradiated samples, suggesting that cosmic rays may induce peptide bond synthesis in glycine crystals. Combining this result with the fact that this compound has the higher solubility among the other glycine polymorphs, we suggest that β-glycine polymorph was relevant for the production of first peptides in the Early Earth. This experimental study helps to understand the survival of such important prebiotic compound which may be delivered by comets/asteroids in the primitive Earth.

The precise amount of interstellar/protostellar glycine in gas phase is still unknown. If glycine is abundant enough in grain/dust surfaces or in asteroids/comets/moons surfaces to become a crystal is still an open question. Besides what could have happened on primitive Earth, higher concentrations of glycine may also have been present after the heavy meteoritic bombardment on others solar system´s bodies such as Mars, Europa, Vesta, and the Moon. Some of these bodies have very rarefied atmospheres (or even none) allowing molecular processing by space ionizing agents such as 1 MeV protons. Nevertheless, we cannot ruled out the investigation of the interaction between such important prebiotic species and space radiation analogs. Future observations and sample extraction from comets or asteroids will help to clarify this issue.

**Acknowledgements**

This work is partially supported by the Brazilian Agencies: CNPq (INEspaço and INCT-A), and FAPERJ. The author thanks Brazilian agency CNPq for research fellowship (#312297/2009-2). The authors also thank the staff of the Van de Graaff Laboratory, Dr. D.P.P Andrade for scientific discussions, Ms. A. Alder-Rangel for English Revision and the Google website for fast and efficient bibliographic access.